\documentclass{icrc2009}

\usepackage{graphicx,units}   

\newcommand{\shorttitle}[1]%
{\markboth{Proceedings of the 31\MakeLowercase{$^{st}$} ICRC, {\L}\'{o}d\'{z} 2009}{#1} }
\newcommand{\etal}{\MakeLowercase{\textit{et al. }}} 

\begin{document}
\title{The Atmospheric Muon Charge Ratio at the MINOS Near Detector.}

\author{\IEEEauthorblockN{J.K. de Jong \IEEEauthorrefmark{1}\IEEEauthorrefmark{2}for the MINOS collaboration.}
                            \\
\IEEEauthorblockA{\IEEEauthorrefmark{1}Illinois Institute of Technology,Chicago Illinois 60616 USA}
\IEEEauthorblockA{\IEEEauthorrefmark{2}now at University of Oxford, Oxford OX1 3RH United Kingdom}}

\shorttitle{J.K. de Jong \etal MINOS charge ratio}
\maketitle

\begin{abstract}
The magnetized MINOS near detector can accurately determine the charge sign of atmospheric muons, this facilitates a measurement of the atmospheric muon charge ratio. To reduce the systematic error associated with geometric bias and acceptance we have combined equal periods of data obtained with opposite magnetic field polarities. We report a charge ratio of $1.2666\pm0.0015(stat.)^{+0.0096}_{-0.0088}(syst.)$ at a mean E$_{\mu,0}cos(\theta)$=\unit[63]{GeV}. This measurement is consistent with the world average\cite{Hebbeker}\cite{Haino}\cite{Achard} but significantly lower than the earlier observation at the MINOS far detector\cite{Adamson}. This increase is shown to be consistent with the hypothesis that a greater fraction of the observed muons arise from kaon decay within the cosmic ray shower. \\
\end{abstract}

 
\section{Introduction}
The MINOS experiment consists of two steel-scintillating sampling calorimeter detectors. The \unit[980]{ton} near detector is used to characterize the spectrum of the neutrino beam and is located at Fermilab in a cavern \unit[94.3]{m} underground at the end of the NuMI beam facility (approximately \unit[1]{km} from the primary proton target). The \unit[5.4]{kton} far detector is located \unit[732]{km} further downstream and is \unit[710]{m} below the surface. MINOS looks for neutrino oscillations by identifying changes to the neutrino spectrum. Both MINOS detectors utilize toroidally magnetized steel planes as the passive absorber material. This magnetic field, which varies between 1 and 2 Tesla, allows them to distinguish between positive and negative muons. \\

When high energy cosmic rays interact with nuclei in the upper atmosphere the subsequent shower produces kaons (K) and pions ($\pi$). These secondary mesons can either interact, or decay to produce muons. Since the majority of cosmic rays are positively charged there will be an excess of positively charged mesons in the shower and consequently an asymmetry in the measured muon charge ratio $N_{\mu^{+}}/N_{\mu^{-}}$. The magnitude of this asymmetry is also influenced by the $\pi$/K production ratio and the fraction of the mesons which decay versus interact. The latter process will be shown to be responsible for the increase in the charge ratio observed by the MINOS far detector\cite{Adamson}\\

\section{The MINOS Near Detector}

 The MINOS near detector\cite{Michael}, located at 88 \ensuremath{^\circ}16' 14" west longitude and 41\ensuremath{^\circ} 50' 26" north latitude on the Fermilab campus, is a steel-scintillator sampling calorimeter with tracking, energy and topology measurement capabilities. It is located \unit[94.3]{m} underground with a flat overburden of 224.6 meters of water equivalent(mwe). It measures \unit[3.8~m~x~4.8~m~x~16.6]{m}. The detector contains 282 vertical steel planes, each \unit[2.5]{cm} thick. Between each plane there is either a \unit[1.0]{cm} thick scintillator plane and a \unit[2.4]{cm} air gap, or a \unit[3.6]{cm} air gap(the scintillator is encased in a \unit[0.1]{cm} thick aluminum skin). Each scintillator plane is comprised of either a full or partial plane which consists of 64 or 96 scintillator strips respectively. The scintillator strips are 4.1 cm wide(transverse direction) and between 2.5 to 4~m in length(longitudinal direction). The first 120 planes located on the upstream portion of the detector comprise the calorimeter. In the calorimeter every 5th plane is fully instrumented covering the cross-sectional area defined by the steel planes. The following 4 planes are partially instrumented. The last 162 planes located on the downstream side of the detector make up the spectrometer. In this region only every 5th plane is instrumented, but instrumented with full scintillator coverage. This region aids in the momentum determination of long tracks. The strips in each scintillator layer are rotated by 90\ensuremath{^\circ} with respect to the previous layer to allow for 3 dimensional track reconstruction. The curvature induced in the track by the magnetic field, which has a strength between 1 and 2 Tesla, coupled with the 3 dimensional track reconstruction allows the determination of the charge sign of the muon track. The detector has two distinct magnetic field polarities. The ``forward'' field focuses $\mu^{{}-{}}$'s coming from the south, toward the center of the detector. The ``reverse'' field focuses $\mu^{{}+{}}$'s coming from the south, toward the center of the detector. \\

\section{Theory}

The differential muon production spectrum which has been parameterized \cite{Gaisser} as 
\begin{eqnarray}
\frac{dN_{\mu}}{dE_{\mu,0}}&\approx&\frac{0.14\cdot E_{\mu,0}^{-2.7}}{cm^{2}sr GeV}\times \\ \nonumber
	& & \left ( \frac{1.0}{1+\frac{1.1\cdot E_{\mu,0}cos(\theta)}{\epsilon_{\pi}}}+\frac{0.054}{1+\frac{1.1E_{\mu,0}cos(\theta)}{\epsilon_{K}}}\right ),
\label{eq:MuonFlux}
\end{eqnarray}
where E$_{\mu,0}$ and $\theta$ are the muons surface energy and zenith angle respectively. The values $\epsilon_{\pi}=$\unit[115]{GeV} and $\epsilon_{K}$=\unit[850]{GeV} are the critical energies above which the pions and kaons would prefer to interact more often than decay. The 0.054 value is related to the $\pi$/K ratio in the showers and the branching fraction of the Kaon to a muon. The first term in the brackets is the muon contribution from pion decay, and the second term is the muon contribution from kaon decay. As a function of E$_{\mu,0}$cos($\theta$) the denominator of the Kaon contribution term drops slower ($\epsilon_{\pi}<\epsilon_{K}$) than the pion term meaning that at large values more of the observed muons are coming from Kaon decays. Since strong interaction production channels lead to a muon charge ratio that is greater for kaon decays than that from pion decays the observed charge ratio should increase\cite{Adamson}. Following the prescription in \cite{Adamson} and defining f$_{\pi^{{}+{}}}$ and f$_{K^{{}+{}}}$ as the fraction of all decaying pions and kaons which decay with a detected $\mu^{{}+{}}$, the atmospheric muon charge ratio $N_{\mu^{{}+{}}}/N_{\mu^{{}-{}}}$ is given in eq. (2):
\begin{eqnarray}	
\frac{N_{\mu^{+}}}{N_{\mu^{-}}}&=&\left ( \frac{f_{\pi^{{}+{}}}}{1+\frac{1.1E_{\mu,0}cos(\theta)}{\epsilon_{\pi}}}+
\frac{0.054f_{K^{{}+{}}}}{1+\frac{1.1E_{\mu,0}cos(\theta)}{\epsilon_{K}}} \right ) \\ \nonumber
	&   &
	 \overline{\left ( \frac{(1-f_{\pi^{{}+{}}})}{1+\frac{1.1E_{\mu,0}cos(\theta)}{\epsilon_{\pi}}}+
\frac{0.054(1-f_{K^{{}+{}}})}{1+\frac{1.1E_{\mu,0}cos(\theta)}{\epsilon_{K}}} \right )}.
\label{eq:CRecostheta}
\end{eqnarray}

\section{The Analysis}

The desired data sample for this analysis consists of atmospheric muons with well reconstructed energy and charge sign. A series of pre-analysis selections are made to reject events which are not consistent with atmospheric muons and to reject events whose origin and kinematics can not be determined. Further selections are applied to ensure track reconstruction quality and that the charge sign of the candidate muon track is accurately determined. This analysis is performed on 190.2 days of data, roughly 450 million triggers, collected between 2006 and 2008. \\

This analysis has been constructed with the intention of reducing the two dominant systematic errors for the charge ratio measurement: geometric bias and randomization. Geometric bias refers to a preferential selection of one charge sign over another, whereas randomization refers to an inability to determine the charge which results in a random charge being assigned to the track. The variables discussed in the following sections were derived using monte-carlo and validated using the data sample. \\


\subsection{Geometric Bias}

\begin{figure}[htb]
\begin{center}
\includegraphics[width=0.4\textwidth]{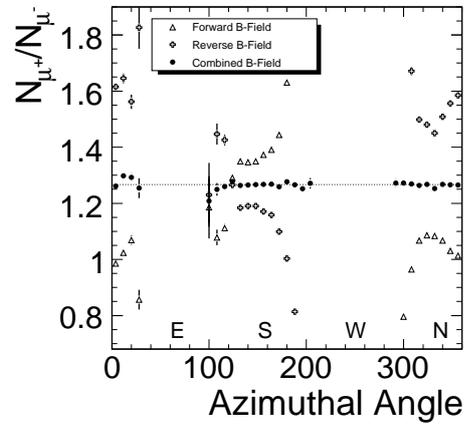}  
\caption{The charge ratio, after all the analysis cuts, varies considerably due to acceptance effects in either field configuration as a function of azimuth angle: Closed circles are the ratios observed during forward field running, and open circles during reversed field running. The combined field has been determined using the geometric mean. The error bars are statistical only. }
\label{fig:CRvsAzimuth}
\end{center}
\end{figure}

Due to east-west asymmetry of the detector (the magnetic coil hole is off-center), as well as the north-south asymmetry introduced by the spectrometer, the detector preferentially reconstructs northerly going muons which occur on the east side of the detector. The forward field focuses northerly going $\mu^{-}$ and defocuses $\mu^{{}+{}}$, therefore the forward field charge ratio will be less than the true charge ratio; similarly, the reversed field charge ratio should be greater for northerly going muons(and vice-versa for southerly going). This is the trend that is observed in figure  \ref{fig:CRvsAzimuth}.

Cancelling the biases due to geometric acceptances has been previously discussed in \cite{Adamson}. If $\epsilon_{1}$ is the acceptance for $\mu^{{}+{}}$ and $\epsilon_{2}$ is the acceptance for $\mu^{{}-{}}$ in the forward field direction(FF) then the acceptance for  $\mu^{{}+{}}$ and  $\mu^{{}-{}}$ in the reverse field direction(RF) are  $\epsilon_{2}$ and $\epsilon_{1}$ respectively. Two independent equations for the charge ratio, r$_{a}$ and r$_{b}$, can be constructed where the acceptance effects cancel. These ratios, corrected for detector lifetime, are: 
\begin{equation}
r_{a}=(N_{\mu^{{}+{}}}/t)_{FF}/(N_{\mu^{{}-{}}}/t)_{RF},
\label{eq:ra}
\end{equation}
and
\begin{equation}
r_{b}=(N_{\mu^{{}+{}}}/t)_{RF}/(N_{\mu^{{}-{}}}/t)_{FF},
\label{eq:rb}
\end{equation}
By combining equations (\ref{eq:ra}) and (\ref{eq:rb}) we obtain a charge ratio in which the geometric acceptance and life-time biases cancel.
\begin{equation}
\frac{N_{\mu^{{}+{}}}}{N_{\mu^{{}-{}}}}=\sqrt{r_{a}r_{b}}=\sqrt{(\frac{N_{\mu^{{}+{}}}}{N_{\mu^{{}-{}}}})_{FF}(\frac{N_{\mu^{{}+{}}}}{N_{\mu^{{}-{}}}})_{RF}}
\label{eq:finaleq}
\end{equation}
The charge ratio in equation (\ref{eq:finaleq}) is also independent of any changes in the muon flux due to seasonal variations that may have occurred between the forward and reversed field running\cite{deJong}. \\ 

\subsection{Reducing Randomization}

Randomization refers to any process which results in the track charge being assigned randomly. In fact any process which degrades ``track quality'' introduces randomization into the data sample. A well reconstructed track is defined as a track whose reconstructed points correspond to strips where a signal was detected. If a ``noise'' hit were to be included as a hit on the track, the reconstruction algorithm will attempt to include this hit in the track and pull the best fit curve away from the true signal hits. Since noise hits are distributed randomly around a track, the curvature induced by this noise hit will result in a random charge being assigned. A well reconstructed track is required to have a maximum deviation from the center of each strip no more than 3~cm in the transverse direction and no more than $1/2$ the strip length plus 1~cm in the longitudinal direction. It is required that at least 60\% of the hits recorded in the detector be part of the track in order to eliminate particularly noisy events. Finally, it's required that the track be reconstructed with a good $\chi^{2}$ per degree of freedom. \\

Track quality ensures that the track is well reconstructed, however when determining the atmospheric muon charge ratio the charge sign of the track needs to be identified accurately as well. The Kalman filter \cite{Fruhwirth} assigns for each track a quantity of $(q/p)\pm\sigma{(q/p)}$, where $q=\pm1$ is the track charge and $p$ the track momentum. A track with a large value of $(q/p)/\sigma(q/p)$ is a track for which the fitter has a high confidence in the assigned charge and momentum. Secondly, a long track occurring in a region of high magnetic field would have a high degree of curvature, and the more hits that occurred along this track the greater the confidence in that curvature. This selection variable will be referred to as the ``{\it BdL}''. The {\it BdL} cut was originally constructed in the far detector analysis\cite{Adamson} to ensure that the magnitude of the bending due to curvature was larger than apparent bending due to multiple scattering. These two selection criteria have been optimized in data, and validated in Monte-Carlo, to maximize selection efficiency and minimize charge sign mis-identification.\\

\begin{figure}[h]
\begin{center}
\includegraphics*[width=0.39\textwidth,angle=0,clip]{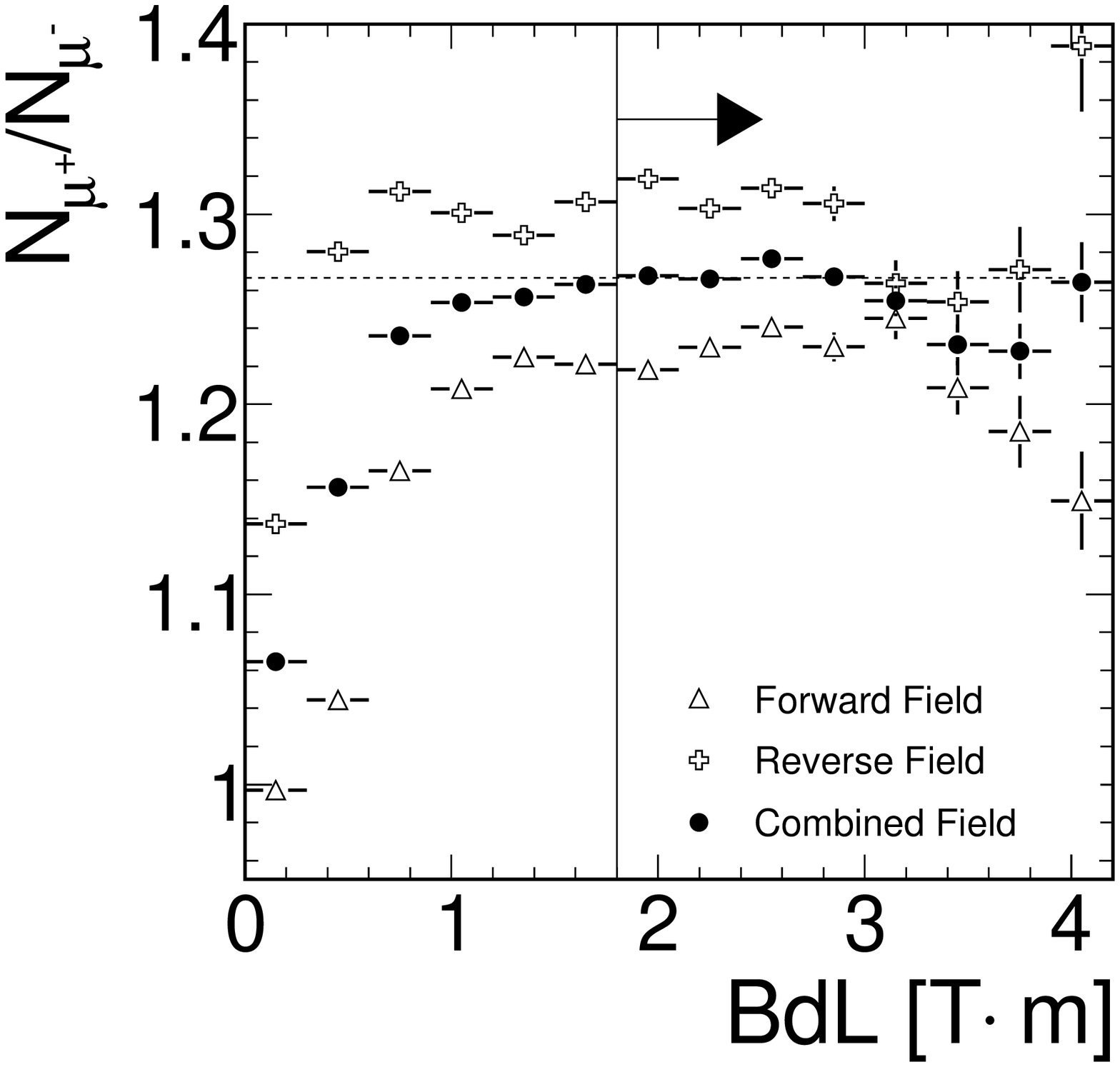}
\includegraphics*[width=0.39\textwidth,angle=0,clip]{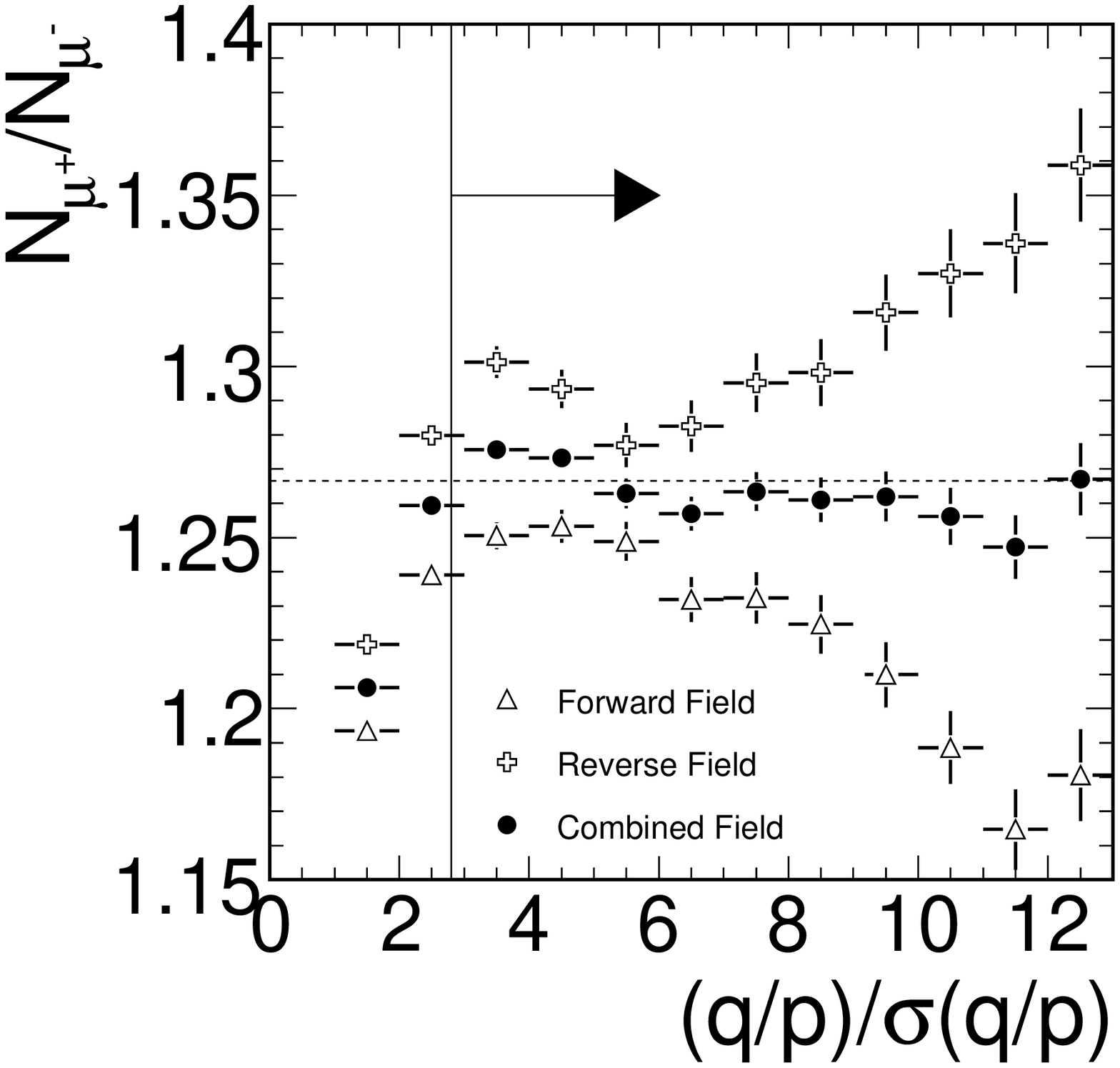}

\caption{\label {fig2} The charge ratio as a function of (t) of {\it BdL}, after all the track quality cuts and a $(q/p)/\sigma(q/p)>$2.8 and (b) as a function of $(q/p)/\sigma(q/p)$ after all cuts and a {\it BdL} cut of 1.8. The vertical lines indicate the values used for the analysis cuts. Error bars are statistical only.}
\end{center}
\end{figure}

Figure \ref{fig2} plots the charge ratio as a function of both $(q/p)/\sigma(q/p)$ (after the curvature cuts and including  $BdL>$1.8) and the minimum plane cut (curvature+$(q/p)/\sigma(q/p)>$2.8). In both cases we see that the charge ratio flattens off indicating that the systematics due to randomization have been reduced. After all the analysis cuts have been applied we retain 2,739,349 events(0.608\% acceptance), with 1,530,719 $\mu^{+}$ and 1,208,530 $\mu^{-}$ for a charge ratio of 1.2666$\pm$0.0015(stat.)

\section{Discussion of Errors}

We consider three specific systematic errors, these are: non-perfect magnetic field inversion, cut-based errors and remnant randomization. \\

We have made the assumption that the systematic error associated with geometric acceptance cancels when determining r$_{a}$ and r$_{b}$ (equations \ref{eq:ra} and \ref{eq:rb} respectively). Had these cancellations been exact we would expect r$_{a}$=r$_{b}$, instead we measure r$_{a}$=1.2625 and r$_{b}$=1.2706. Half the difference between these two values, $\pm$0.0041 is taken as the systematic uncertainty associated with the magnetic field inversion. \\

The largest systematic uncertainty comes from the structure observed in the plot of charge ratio versus (q/p)/$\sigma$(q/p) in Fig.~\ref{fig2}. Increasing the strength of the applied (q/p)/$\sigma$(q/p) cut increases the charge ratio(decreases the randomization) and reaches a maximum value at $(q/p)/\sigma(q/p)\approx$3.5, it then falls slightly and flattens off for $(q/p)/\sigma(q/p)>5.0$. Classifying this bump as a randomization error would imply, counter-intuitively, that the charge identification is worse at large values(low momentum) of  $(q/p)/\sigma(q/p)$ than at smaller values(high momentum). The charge ratio was calculated for the peak region 2.8$<(q/p)/\sigma(q/p)<$5.0 as 1.27444, and for the flat region of the graph $(q/p)/\sigma(q/p)>$5.0 as 1.260. The maximum deviation from the average charge ratio is 0.0078, which is taken as an uncertainty. \\

Monte Carlo studies suggest that the analysis miss-identifies the charge on 0.2555\% of the events, which implies the true charge ratio should be 1.2682 (a change of +0.0016). However, given the complexity of the detector it is expected that the true miss-identification rate will be larger in the data than in the Monte Carlo simulation. Further data to monte-carlo comparison studies indicate that the charge miss-identification in data is 2.3 times higher than in monte-carlo. The result is a one-sided systematic of +0.0037. \\

After all analytical selections the charge ratio measured at the MINOS near detector is:
\begin{equation}
R= 1.2666\pm0.0015(stat.)^{+0.0096}_{-0.0088}(syst.). \\
\end{equation}

\section{Atmospheric Muon Charge Ratio}

Using the technique outlined in \cite{Juergen} the energy lost $E_{loss}$ by muons traversing the near detector overburden has been calculated as a function of reconstructed track momentum $E_{\mu,det}$ and zenith angle $\theta$ . For this analysis we take the Near Detector cavern floor to be situated at 123.8 meters above sea level(msl), and the roof at 133.5 msl. Directly above the cavern hall lies 72.1 m of Dolomite/Shale bedrock ( CaMg($C0_{3}$)$_{2}$ w/ 8.0\% ~H$_{2}$O by weight) at a density of 2.41 g/cm$^{3}$, followed by 22.2 meters of Glaciel till with a density of 2.29 g/cm$^{3}$. This gives a flat vertical overburden of 224.6 meters of water equivalent(mwe). The energy lost by muons traversing the overburden increases with both detected momentum and zenith angle. The muon energy at the surface is simply:
\begin{equation}
E_{\mu,0}=E_{\mu,det}+E_{loss}(E_{\mu,det},\theta).
\label{eq:addition}
\end{equation}

\begin{figure}[htb]
\begin{center}
\includegraphics[width=0.4\textwidth]{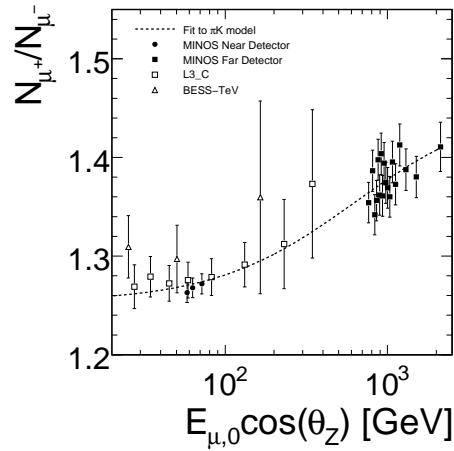}
\caption{Charge ratio as a function of E$_{\mu,0}$cos($\theta$). The MINOS near detector agrees very well with the L3+C measurements. The dashed line is the best fit value to the $\pi$K model.}
\label{fig:CRecos}
\end{center}
\end{figure}

The requirement that the muon track possesses a charge confidence value (q/p)/$\sigma$(q/p)$>$2.8, is a statement that the track must bend significantly in the magnetic field. The detector's magnetic field and spatial resolution limits the maximum detectable momentum for a muon in this analysis. Since MINOS is a shallow underground detector with a low maximum detectable momentum, the E$_{\mu,0}$cos($\theta_{Z}$) distribution is narrow only about 30~GeV. The MINOS near detector measurement is observed to be consistent with previous experiments and lower than the TeV scale measurement from the Far Detector. A $\chi^{2}/ndof$ fit test to the $\pi$K model, equation (2), was performed over f$_{K}$/f$_{\pi}$ space using the L3+C, MINOS near and far detector charge ratio results. A $\chi^{2}/ndof$ minimum of$~$0.474 is found at f$_{\pi}$=0.5488$\pm$0.0016 and f$_{K}$=0.7021$\pm$0.011. The best fit curve to the data is plotted in Fig.~\ref{fig:CRecos}. 

\label{eq:CRecostheta}

\section{Conclusion}

The atmospheric muon charge ratio has been measured using 190.2 days of data at the MINOS near detector to be 1.2666 $\pm$ 0.0015(stat.)$^{+0.0096}_{-0.0088}$ (syst.). The $\pi$K model was shown to fit both the near and far detector data implying that the observed increase in charge ratio at the far detector is dominated by the increased fraction of observed muons arising from kaon decay. The MINOS collaboration has collected 120 more days of data which will be added to this dataset and published at a later date.



  
\section{Acknowledgements}  
This work was supported by the US DOE, the UK STFC, the US NSF, the State and University of Minnesota, the University of Athens, Greece and Brazil's FAPESP and CNPq. We are grateful to the Minnesota Department of Natural Resources, the crew of Soudan Underground Laboratory, and the staff of Fermilab for their contributions to this effort.

\end{document}